\documentclass[conference]{IEEEtran}
\usepackage{etoolbox}
\usepackage{amsmath,amssymb,amsfonts}
\usepackage{algorithm}
\usepackage{algpseudocode}
\usepackage{bbding}
\usepackage{float} % 在导言区引入 float 宏包
\usepackage{graphicx}
\usepackage{textcomp}
\usepackage{wrapfig}
\usepackage{stfloats}
\usepackage{placeins}
\usepackage{booktabs}
\usepackage{tabularray}
\usepackage{multirow}
\usepackage[table,xcdraw]{xcolor}  % 导入颜色包
\ifCLASSOPTIONcompsoc
\usepackage[caption=false, font=normalsize, labelfont=sf, textfont=sf]{subfig}
\else
\usepackage[caption=false, font=footnotesize]{subfig}
\fi
\usepackage{bbding} %首先在导言区调用bbding包
\usepackage{cite}    
\makeatletter
\let\NAT@parse\undefined
\makeatother
\usepackage{hyperref}
\hypersetup{
    colorlinks=true,            % 激活链接颜色，去掉链接边框
    linkcolor=blue,              % 文档内部链接颜色（如图表等引用）
    citecolor=blue,            % 文献引用链接颜色  % 文件链接颜色
    urlcolor=blue         % 外部URL链接颜色
}
\IEEEoverridecommandlockouts
\usepackage{cite}
\usepackage{amsmath,amssymb,amsfonts}
\usepackage{graphicx}
\usepackage{textcomp}
\usepackage{xcolor}
\def\BibTeX{{\rm B\kern-.05em{\sc i\kern-.025em b}\kern-.08em
    T\kern-.1667em\lower.7ex\hbox{E}\kern-.125emX}}
\begin{document}

\title{DT4ECG: A Dual-Task Learning Framework for ECG-Based Human Identity Recognition and Human Activity Detection\\

}

\author{
\IEEEauthorblockN{
Siyu You\textsuperscript{1}, Boyuan Gu\textsuperscript{1}, Yanhui Yang\textsuperscript{1}, Shiyu Yu\textsuperscript{1}, Shisheng Guo\textsuperscript{2}\textsuperscript{*}}
\IEEEauthorblockA{\textsuperscript{1} Glasgow College, University of Electronic Science and Technology of China, Chengdu, China\\
\textsuperscript{2} School of Information and Communication Engineering, University of Electronic Science and Technology of China, Chengdu, China}}

\maketitle

\begin{abstract}
This article introduces DT4ECG, an innovative dual-task learning framework for Electrocardiogram (ECG)-based human identity recognition and activity detection. The framework employs a robust one-dimensional convolutional neural network (1D-CNN) backbone integrated with residual blocks to extract discriminative ECG features. To enhance feature representation, we propose a novel Sequence Channel Attention (SCA) mechanism, which combines channel-wise and sequential context attention to prioritize informative features across both temporal and channel dimensions. Furthermore, to address gradient imbalance in multi-task learning, we integrate GradNorm, a technique that dynamically adjusts loss weights based on gradient magnitudes, ensuring balanced training across tasks. Experimental results demonstrate the superior performance of our model, achieving accuracy rates of 99.12\% in ID classification and 90.11\% in activity classification. These findings underscore the potential of the DT4ECG framework in enhancing security and user experience across various applications such as fitness monitoring and personalized healthcare, thereby presenting a transformative approach to integrating ECG-based biometrics in everyday technologies.
\end{abstract}

\section{Introduction}
\label{sec:introduction}
\IEEEPARstart{E}{lectrocardiogram} (ECG) is a medical test that records the electrical activity of the heart over time. It is widely used in clinical settings for diagnosing heart diseases, detecting arrhythmias, and monitoring the overall health of individuals \cite{MORENOSANCHEZ}. An ECG signal captures the electrical impulses that trigger each heartbeat and is often considered a unique biometric feature, making it valuable not only for medical purposes but also for identity verification \cite{asadianfam_ecg-based_2024}. Each individual's heart rhythm exhibits distinct characteristics, and this uniqueness has led to the exploration of ECG for biometric identification in recent research \cite{Madhura}, \cite{labati_deep-ecg_2019}. 

Substantial progress has been made in ECG-based biometric identification. Early works focused on extracting time-domain and frequency-domain features from ECG signals, using classical machine learning techniques such as support vector machines (SVMs) and k-nearest neighbors (K-NN) for person identification \cite{Mohamed}. More recently, with the advancement of deep learning, methods such as convolutional neural networks (CNNs) \cite{Cho_CNN},\cite{wu_cnn} and recurrent neural networks (RNNs) \cite{salloum_RNN} have been explored to improve the accuracy and robustness of ECG-based identity recognition systems. These deep learning models are capable of automatically learning discriminative features directly from the raw ECG signals, enabling them to outperform traditional methods.

In addition to identity recognition, human activity recognition (HAR) has become a critical task in understanding and monitoring individual health status, typically involving the classification of activities such as resting, active (during exercise or movement), and post-exercise recovery. These activities provide valuable insights into an individual’s health and fitness level, enabling more tailored and effective interventions. ECG signals, capturing unique physiological changes in heart rate and variability patterns, offer a distinct perspective for HAR by detecting subtle responses to physical activity. Currently, numerous ECG-based HAR methods have been developed. For instance, Arani \textit{et al.} integrated ECG signals with 3D-ACC signals, applied hand-crafted time and frequency domain feature extraction, and utilized random forest models for HAR \cite{Arani_ECG_HAR}. Compared with such methods that rely on manual feature selection, approaches based on CNNs have gained popularity for their ability to automatically extract features. Yun \textit{et al.} developed an ECG-based HAR system using a one-dimensional convolutional neural network (1D-CNN) to analyze wireless ECG data from 40 participants performing five daily activities. Their method achieved an overall test accuracy of 82.9\%, with particularly high accuracy in recognizing the sleeping activity at 98.5\% \cite{Yun_HAR}.

Despite the significant progress in both fields, ECG-based identity recognition and activity classification are generally treated as separate tasks. Few studies have attempted to combine these two tasks into a multi-task learning (MTL) framework, which could benefit from the shared representation learning between the two. MTL has been shown to improve performance by leveraging common features across tasks and enabling the model to generalize better on each individual task \cite{mtl_review}. For instance, in the context of health monitoring, jointly predicting a person’s identity and their current activity could help in better personalizing the system’s response and provide more comprehensive real-time health assessments. The combination of ECG-based identity recognition and activity classification offers significant potential in creating a more comprehensive and robust health monitoring system. By integrating both tasks, a system can simultaneously authenticate users while tracking their activity status, improving both security and user experience in various applications such as fitness monitoring, elderly care, and personalized health systems.

In this paper, we propose a novel MTL framework, DT4ECG, that simultaneously handles person identification and human activity detection tasks based on ECG signals. The main contributions of this work can be summarized as follows:

1) We introduce a MTL model designed to simultaneously perform person classification and activity classification from ECG signals. The model leverages a 1D-CNN as backbone to extract features and integrates residual blocks to enhance its feature extracting ability.

2) We propose a novel attention mechanism called the Sequence Channel Attention (SCA) mechanism, which combines channel-wise attention and sequential context attention. This mechanism allows the model to focus on the most informative features from both the channel and temporal dimensions of the ECG signal.

3) We integrate the GradNorm technique to address the issue of imbalance task gradients and balance the gradients during training. GradNorm dynamically adjusts the loss weights for each task based on their gradient magnitudes, ensuring that both person classification and activity classification tasks contribute equally to the model's learning process.

The rest of the paper is organized as follows: Section \ref{sec::Experimental setup} presents the experimental setup, including the dataset and evaluation metrics. Section \ref{sec::methodology} describes the methodology, detailing the MTL framework and the attention mechanism used. Section \ref{sec::results} presents the experimental results and discussions. Finally, Section \ref{sec::conclusion} concludes the paper and outlines future work.
\section{Experimental Setup}
\label{sec::Experimental setup}

\subsection{Data Acquisition}
In this article, we did not use any open-source ECG datasets, such as MIT-BIH \cite{MITBIH} or ECG-ID \cite{ecgid}, because these datasets do not specifically include data collected under different physical activity, which are critical for our study. Therefore, we chose to collect a custom self-built dataset that includes ECG signals from subjects under varying movement conditions, ensuring that our data more accurately represents real-world scenarios involving different activity levels.

Fifteen healthy subjects, aged between 20 and 29 years, voluntarily participated in the experiment. The experimental protocol was reviewed and approved by experts to ensure minimal risk to the participants.

\begin{figure}
    \centering
    \includegraphics[width=\linewidth]{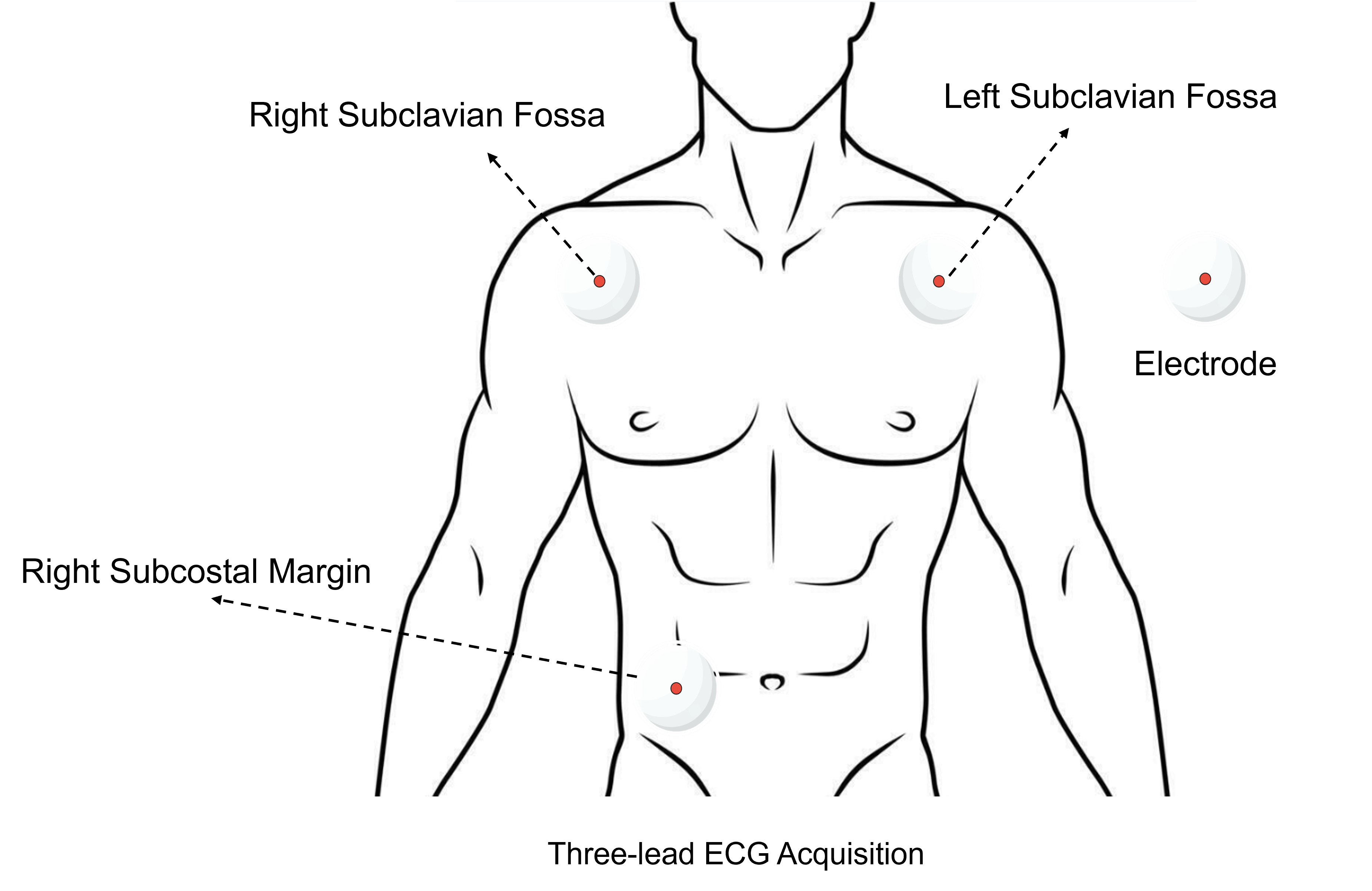}
    \caption{The three-lead electrodes method to capture ECG signal}
    \label{fig:3leads}
\end{figure}

The ECG signals were recorded using a three-lead electrode configuration, as shown in Fig. \ref{fig:3leads}. This method offers both convenience and high accuracy, making it suitable for reliable heartbeat wave acquisition. The three electrodes were placed at the right subclavian fossa, left subclavian fossa, and the right subcostal margin, ensuring that the heart was positioned centrally for accurate monitoring.

\begin{figure}
    \centering
    \includegraphics[width=\linewidth]{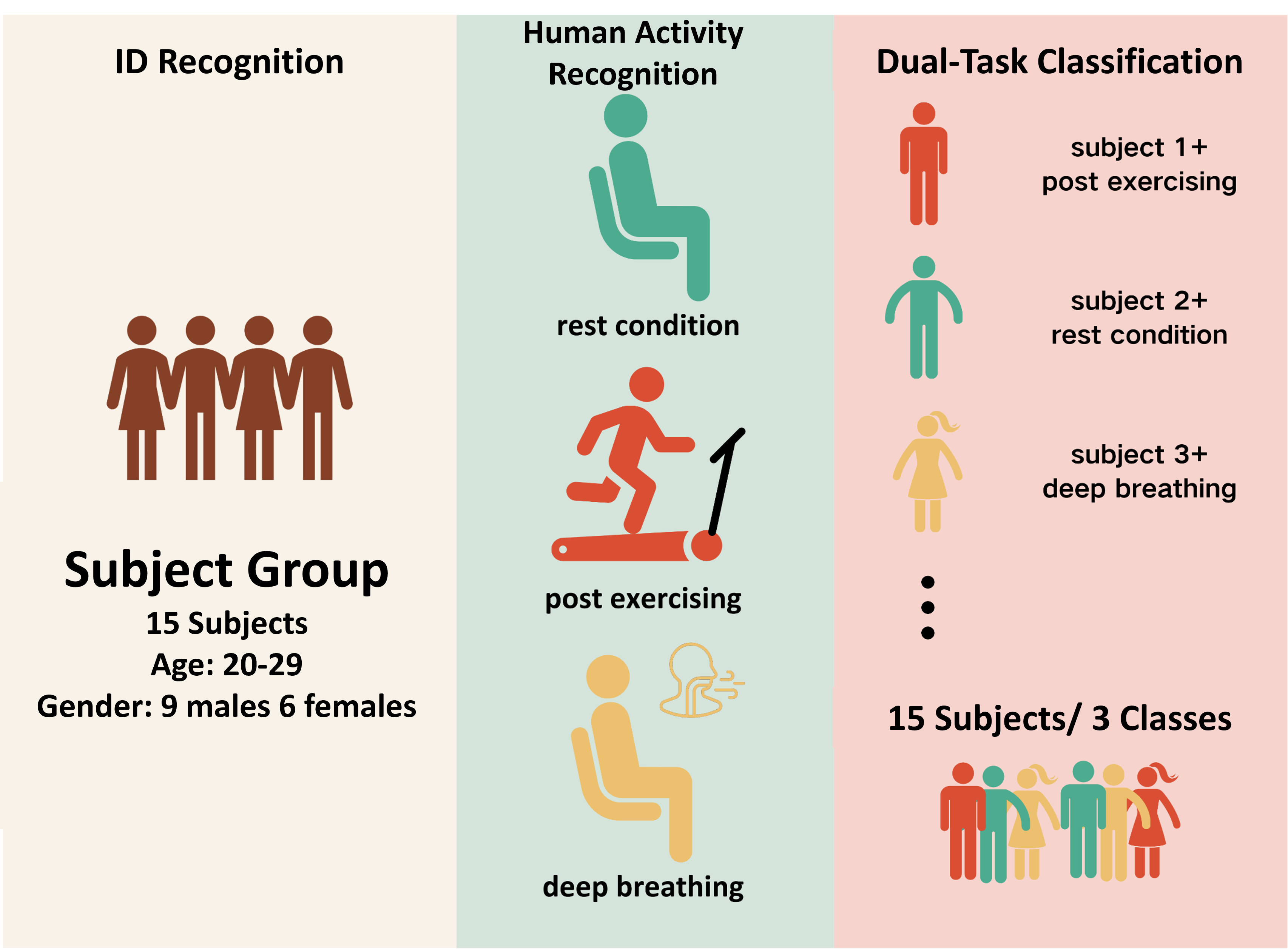}
    \caption{The configurations of subject group}
    \label{fig:subjectgroup}
\end{figure}

For each subject, three sets of data, each lasting 3 minutes, were collected under the following conditions: resting, exercising, and deep-breathing. To maintain consistency across all subjects, the exercise protocol involved running at a speed of 9 kilometers per hour for 3 minutes, followed by data collection 1 minute after the exercise. The deep-breathing condition consisted of a breathing pattern with a 5-second inhalation and exhalation cycle. The subject group configuration is depicted in Fig. \ref{fig:subjectgroup}.

\subsection{Data Preprocessing}

The ECG data acquired using the three-lead electrode configuration was recorded through the BIOPAC MP 160 system. The sampling frequency is 100 Hz.

To enhance the quality of the ECG signals, preprocessing was conducted using MATLAB 2022b. The preprocessing steps are outlined as follows:

\begin{itemize}
    
\item \textbf{Filtering:} A high-pass filter with a cutoff frequency of 0.5 Hz and a notch filter at 60 Hz were applied to remove primary noise in the ECG signal, such as baseline wander and power-line interference, thereby preserving the cardiac signal.

\item \textbf{Data Normalization:} The ECG signal amplitudes were normalized to a range of 0 to 1 to facilitate subsequent modeling.

\item \textbf{Signal Segmentation:} The continuous ECG signal was segmented into 3-second slices, with each slice containing 300 data points, to prepare the signal for model input.
\end{itemize}

\section{Methodology}
\label{sec::methodology}
\subsection{1D Residual Module}
\begin{figure}[h]
    \centering
    \includegraphics[width=0.5\linewidth]{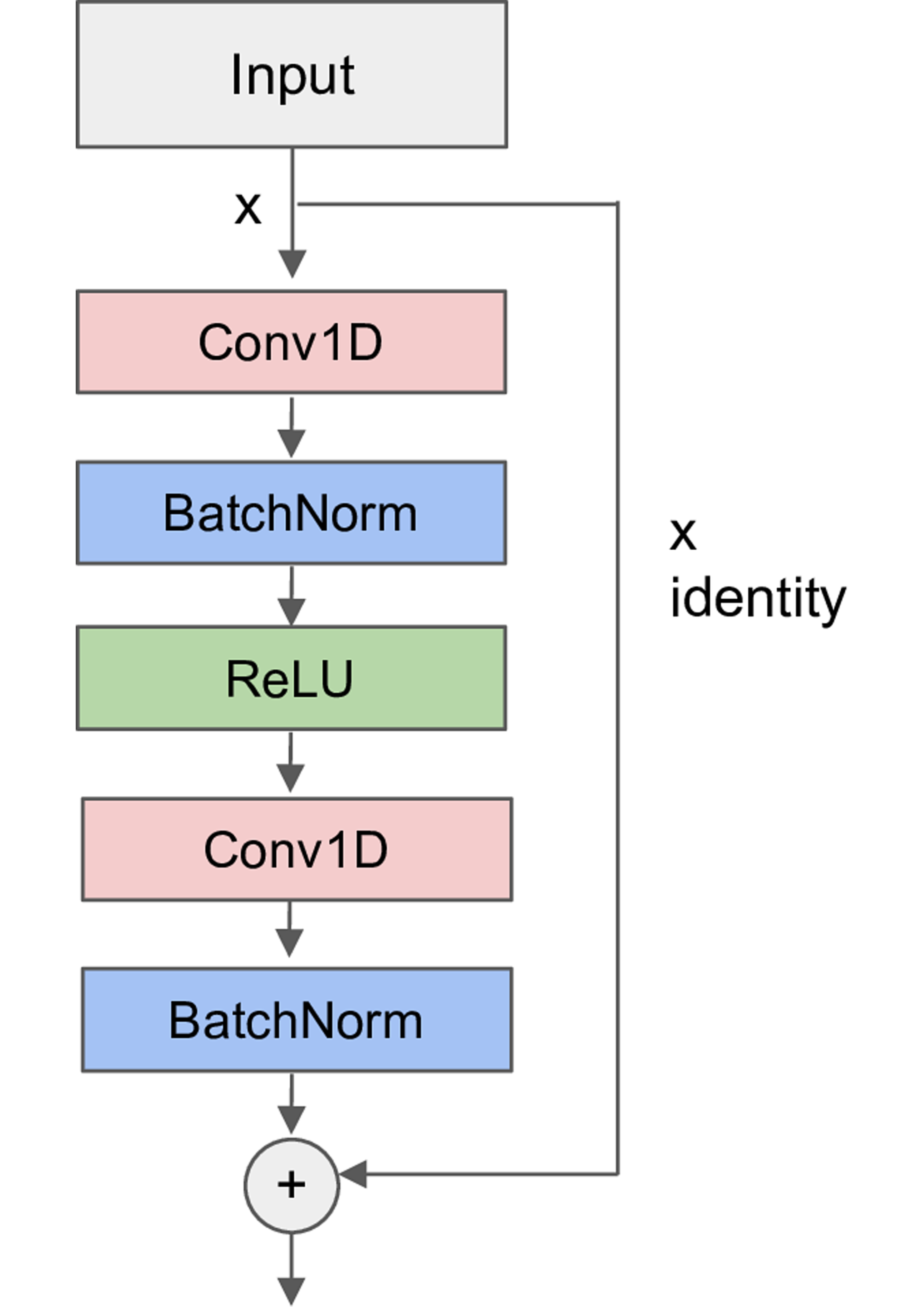}
    \caption{The overview of 1D residual module}
    \label{fig:1dres}
\end{figure}

In the proposed method, we utilize 1D residual module to capture complex features in the ECG signals. This module is derived from classic residual networks (ResNet) used in computer vision tasks proposed by He \textit{et al.}\cite{ResNet}. The core idea behind the residual connection is to facilitate the learning of residual mappings rather than directly learning the desired output, helping to overcome the vanishing gradient problem in deep networks. Fig. \ref{fig:1dres} demonstrates the 1D residual module used in our network. This module consists of two 1D convolutional layers, each followed by batch normalization and ReLU activation. The first convolutional layer captures low-level features, while the second layer refines these features. The residual connection enables the direct addition of the input signal to the output of the second convolution, allowing the model to focus on learning residual mappings instead of the entire transformation. This design helps to mitigate the issue of vanishing gradients and facilitates the training of deeper models.

\subsection{SCA Mechanism}

\begin{figure}[h]
    \centering
    \includegraphics[width=\columnwidth]{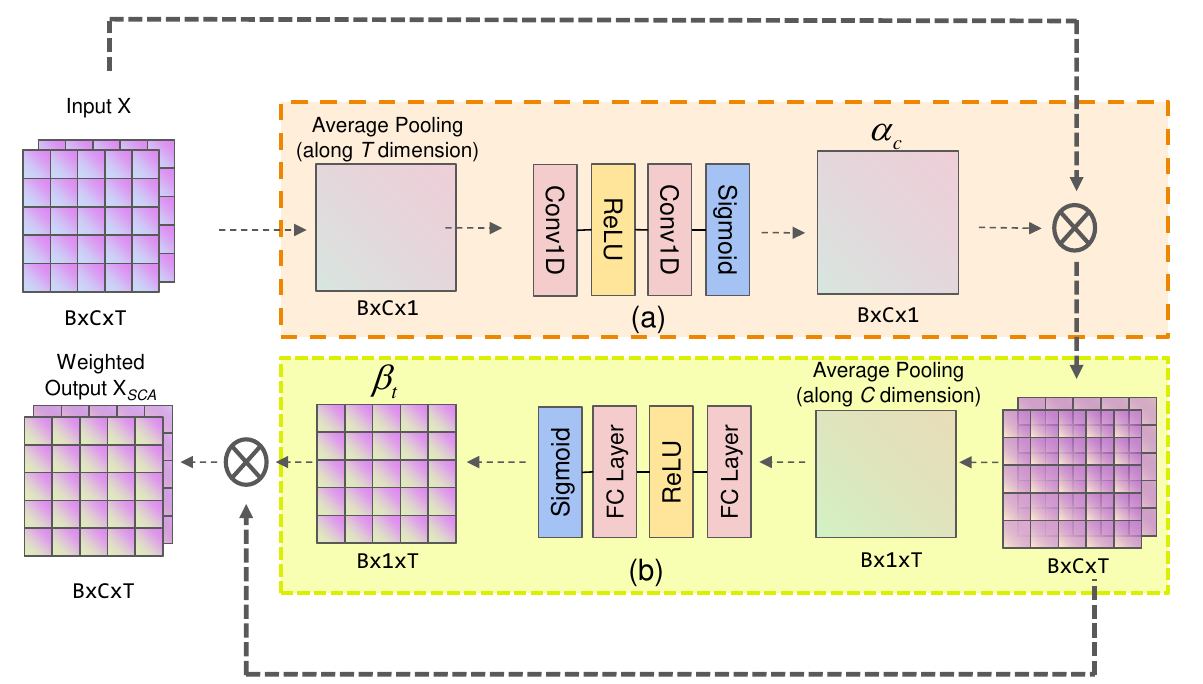}
    \caption{The workflow of the SCA module. (a) Channel attention mechanism (b) Sequence attention mechanism}
    \label{fig:SCA}
    \vspace{-1em}
\end{figure}

Conventional 1D-CNNs are effective at capturing local feature representations in sequential signals. However, these models often fail to fully exploit the temporal dependencies and global context of the signals, which are crucial for tasks such as signal classification or time-series analysis. To address this issue, we propose a novel SCA mechanism that enables the model to focus on more important features, both spatially (in the channel dimension) and temporally (across the time axis). The SCA mechanism combines channel attention and sequence attention in a unified manner and its workflow is shown in Fig. \ref{fig:SCA}.

\subsubsection{Channel Attention Mechanism}

The channel attention mechanism focuses on learning which channels of the input feature map are most important for the task. It leverages global feature dependencies by performing pooling operations along the temporal dimension, enabling the model to selectively highlight the most informative channels.

In the SCA mechanism, we first apply average pooling to the input feature map \( \mathbf{X} \in \mathbb{R}^{B \times C \times T} \), where \( B \) is the batch size, \( C \) is the number of channels, and \( T \) is the number of time steps. The pooling operation outputs a tensor of size \( \mathbf{X}_{avg} \in \mathbb{R}^{B \times C \times 1} \), which aggregates the information from each channel across all time steps. To learn the importance of each channel, this pooled feature is passed through a series of convolutional layers:

\begin{equation}
\mathbf{z}_c = Conv1d({AvgPool}(\mathbf{X})),
\end{equation}
where \( \mathbf{z}_c \in \mathbb{R}^{B \times C \times 1} \) represents the output of the convolutional layers. The channel attention weight is computed by applying the sigmoid activation function:

\begin{equation}
\boldsymbol{\alpha}_c = \sigma(\mathbf{z}_c),
\end{equation}
where \( \boldsymbol{\alpha}_c \in \mathbb{R}^{B \times C \times 1} \) represents the attention weight for each channel. Finally, the attention weights are applied to the input feature map \( \mathbf{X} \) through element-wise multiplication:

\begin{equation}
\mathbf{X'} = \mathbf{X} \cdot \boldsymbol{\alpha}_c.
\end{equation}

\subsubsection{Sequence Attention Mechanism}

The sequence attention mechanism focuses on learning which time steps of the input feature map are most important for the task. This mechanism operates by pooling features across all channels for each time step, thereby learning which time steps contain the most relevant information for the given task.

To compute sequence attention, we first apply average pooling across the channel dimension for each time step of the feature map \( \mathbf{X'} \in \mathbb{R}^{B \times C \times T} \), producing a tensor of size \( \mathbf{X'}_{avg} \in \mathbb{R}^{B \times 1 \times T} \). This pooled feature is passed through a fully connected (FC) layer to compute the attention scores for each time step:

\begin{equation}
\mathbf{s}_t = FC(AvgPool(\mathbf{X'})),
\end{equation}
where \( \mathbf{s}_t \in \mathbb{R}^{B \times 1 \times T} \) is the output of the fully connected layer. The attention weights for each time step are then computed using the sigmoid activation function:

\begin{equation}
\boldsymbol{\beta}_t = \sigma(\mathbf{s}_t),
\end{equation}
where \( \boldsymbol{\beta}_t \in \mathbb{R}^{B \times 1 \times T} \) represents the sequence attention weights. Finally, the sequence attention is applied to the feature map \( \mathbf{X'} \) via element-wise multiplication:

\begin{equation}
\mathbf{X''} = \mathbf{X'} \cdot \boldsymbol{\beta}_t.
\end{equation}

\subsubsection{Final Output of the SCA Module}

The final output of the SCA module combines both channel attention and sequence attention. The feature map after applying both attention mechanisms is obtained by:

\begin{equation}
\mathbf{X}_{SCA} = \mathbf{X} \cdot \boldsymbol{\alpha}_c \cdot \boldsymbol{\beta}_t,
\end{equation}
where \( \mathbf{X}_{SCA} \in \mathbb{R}^{B \times C \times T} \) is the output of the SCA module, which is the result of applying both the channel and sequence attention mechanisms to the input feature map \( \mathbf{X} \).

By integrating both types of attention, the SCA module enables the model to focus on more informative features, both spatially (across channels) and temporally (across time steps). This leads to improved feature representations, making it particularly effective for tasks that involve sequential data such as signal classification and time-series forecasting.

\subsection{GradNorm Loss}

In MTL, one common challenge is the disparity in the convergence rates of different tasks. While some tasks may converge quickly, others might take significantly longer to reach convergence. This happens because tasks often have different levels of difficulty, and their loss functions may exhibit varying magnitudes, which can cause imbalanced gradients during the backpropagation process. If losses from all tasks are simply added without considering their relative difficulties, tasks that converge faster will dominate the gradient updates. As a result, the tasks that are slower to converge might not receive sufficient gradient updates during the early stages of training, leading to inefficient learning and wasted computational resources. To address this issue, we use GradNorm mechanism, which dynamically adjusts the loss scaling during training to ensure that tasks with slower convergence are prioritized \cite{gradnorm}. 
\begin{algorithm}[htbp]
\caption{GradNorm for MTL}
\begin{algorithmic}[1]
\State \textbf{Initialize:} Model parameters, task-specific weights \( w_i \) for each task \( i \), and hyperparameter \( \alpha \)
\For{each training step}
    \State Compute the loss for each task \( \mathcal{L}_i \) for task \( i \)
    \State Calculate the gradient for each task: \( \nabla \mathcal{L}_i \)
    \State Compute the norm of the gradient for each task: \( g_i = \|\nabla \mathcal{L}_i\| \)
    \State Normalize the gradient norms and compute the loss scaling factor:
    \[
    \Delta w_i = \frac{g_i}{\sum_{j=1}^{M} g_j}
    \]
    \State Update the task-specific weight for each task:
    \[
    w_i \leftarrow w_i \cdot \exp\left(\alpha \cdot (\Delta w_i - 1)\right)
    \]
    where \( \alpha \) controls the rate of adjustment.
    \State Compute the weighted total loss:
    \[
    \mathcal{L} = \sum_{i=1}^{M} w_i \cdot \mathcal{L}_i
    \]
    \State Update the model parameters based on the weighted total loss and gradients: 
    \[
    \nabla \mathcal{L}_{\text{total}} = \sum_{i=1}^{M} w_i \cdot \nabla \mathcal{L}_i
    \]
    \State Perform a gradient update step on the model parameters using \( \nabla \mathcal{L}_{\text{total}} \)
\EndFor
\end{algorithmic}
\end{algorithm}
\subsubsection{Cross-Entropy Loss}

In our framework, we use cross-entropy loss for both tasks, as it is commonly used in classification problems. The cross-entropy loss for task \( i \) is defined as:

\begin{equation}
\mathcal{L}_i = -\frac{1}{N} \sum_{n=1}^{N} \left[ y_{i,n} \log(p_{i,n}) + (1 - y_{i,n}) \log(1 - p_{i,n}) \right],
\end{equation}
where \( y_{i,n} \) represents the ground truth label for the \( n \)-th sample in task \( i \), \( p_{i,n} \) is the predicted probability for the \( n \)-th sample, and \( N \) is the total number of samples.

The overall loss function \( \mathcal{L} \) is the weighted sum of the individual losses for each task:

\begin{equation}
\mathcal{L} = \sum_{i=1}^{M} w_i \mathcal{L}_i,
\end{equation}
where \( M \) is the number of tasks, and \( w_i \) is the weight for task \( i \).
\begin{figure*}[ht]
    \centering
    \includegraphics[width=\linewidth]{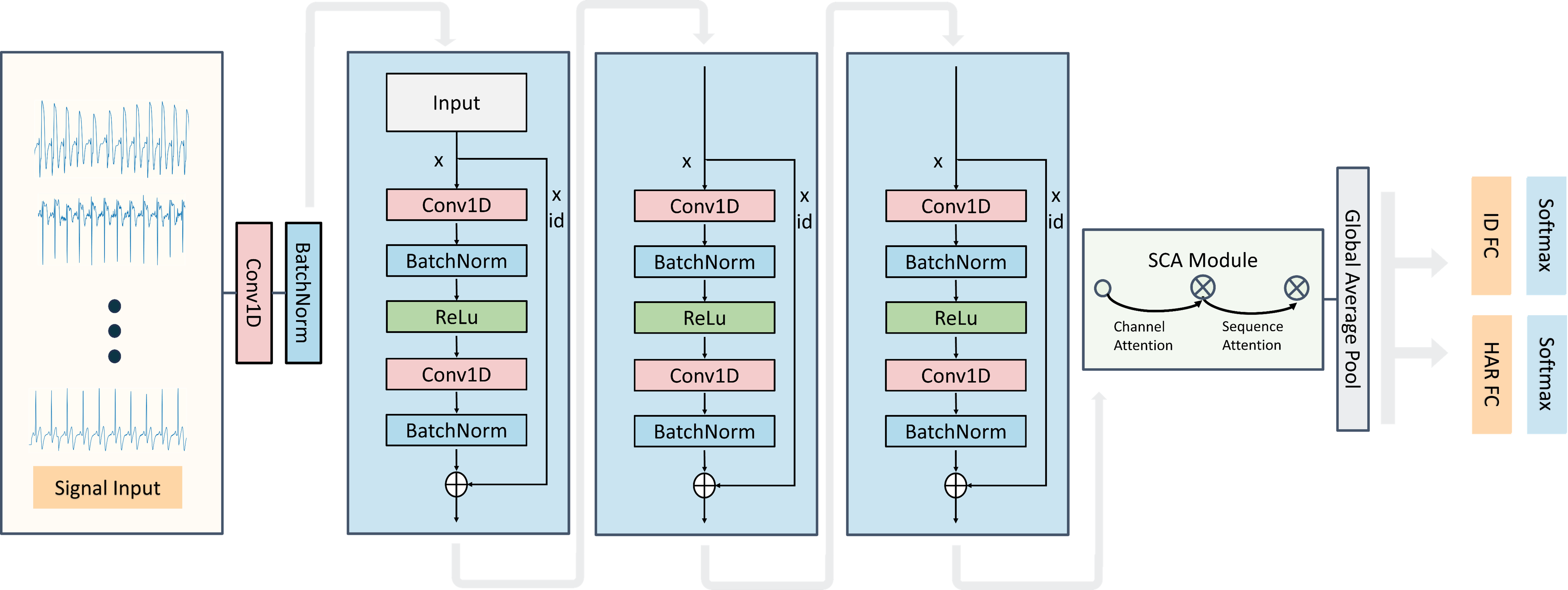}
    \caption{The overview of the proposed DT4ECG framework}
    \label{fig:overview}
\end{figure*}
\subsubsection{GradNorm for Multi-Task Training}
GradNorm aims to equalize the magnitude of gradients across all tasks by adjusting their respective loss weights. The key idea behind GradNorm is to scale the loss for each task according to the relative difficulty of the task, which is measured by the rate of change of the gradients during training.

In this algorithm, the gradient norm \( g_i \) is computed for each task \( i \) by calculating the Euclidean norm of the gradients \( \nabla \mathcal{L}_i \). Then, we compute the loss scaling factor \( \Delta w_i \), which indicates how much the task weights should be adjusted based on the relative gradients across tasks. The weights \( w_i \) are updated by multiplying them with an exponential term, where \( \alpha \) is a hyperparameter that controls the speed of adjustment.

\subsection{The Overview of Model}
The overview of the proposed DT4ECG framework is demonstrated in Fig. \ref{fig:overview}. The initial layers of the model comprise a series of one-dimensional convolutional layers and basic residual blocks. These layers are tasked with feature extraction from the raw time-series data. The architecture begins with an initial convolution layer followed by three 1D residual modules, which help in capturing temporal patterns and local dependencies effectively as the network depth increases. Following the convolution layers, the proposed SCA module is employed to enhance feature representation in both channel and sequence dimensions. After the attention refinement through the SCA module, the features are processed via global average pooling. This technique aggregates the sequence dimension into a compact representation by computing the mean of each feature map. This aggregation step reduces the dimensionality and highlights the most salient information from the input sequence, simplifying the input for subsequent FC layers. The pooled output is then fed into task-specific FC layers, which separately predict outcomes for individual identification and activity classification tasks. Each of these FC layers is followed by a Softmax activation function, which transforms the computed scores into probability distributions across the classes. This enables the model to output the class with the highest probability as the predicted result for each task.

\section{Results and Discussions}
\label{sec::results}
In this section, we compare the performance of the proposed model with baseline model and existing approaches.
\subsection{Training Parameters}

% Please add the following required packages to your document preamble:
% \usepackage{booktabs}
\begin{table*}[]
\caption{Performance comparison of different models. The best performing metrics are highlighted in bold.}
\label{tab:performance_comparison}
\begin{tabular}{ccccc|cccc}
\hline
\multirow{2}{*}{Model} & \multicolumn{4}{c|}{ID (15 classes)}                          & \multicolumn{4}{c}{Activity (3 classes)}                             \\ \cline{2-9} 
                       & Accuracy (\%)  & Precision (\&) & Recall (\%)    & F1-score(\%)   & Accuracy (\%)  & Precision (\%) & Recall (\%)    & F1-score (\%)  \\ \hline
LSTM                   & 36.42          & 36.49          & 36.42          & 35.80          & 39.17          & 39.37          & 39.17          & 39.15          \\
MLP                    & 30.54          & 31.01          & 30.52          & 30.28          & 37.55          & 37.67          & 37.55          & 37.50          \\
1D-CNN                 & 98.12          & 98.16          & 98.12          & 98.11          & 71.46          & 72.57          & 71.46          & 70.93          \\
1D-MobileNet           & 99.00          & 99.02          & 99.00          & 98.99          & 79.22          & 79.05          & 79.22          & 79.09          \\
1D-EfficientNet        & 98.75          & 98.76          & 98.75          & 98.74          & 82.10          & 83.50          & 82.10          & 81.78          \\
1D-ResNet              & 98.37          & 98.39          & 98.37          & 98.36          & 83.98          & 84.16          & 83.98          & 84.01          \\
1D-DenseNet            & 99.00          & 99.01          & 99.00          & 98.99          & 79.72          & 82.13          & 79.72          & 79.85          \\
Proposed Method        & \textbf{99.12} & \textbf{99.14} & \textbf{99.12} & \textbf{99.12} & \textbf{90.11} & \textbf{90.25} & \textbf{90.11} & \textbf{90.11} \\ \hline
\end{tabular}
\end{table*}

% \begin{table*}
% \caption{Descriptions and Results of the ECG-based ID Recognition and HAR of the State-of-the-art Works}
% \label{tab:performance_comparison_SOTA}
% \centering
% \resizebox{0.9\linewidth}{!}{%

% \renewcommand{\arraystretch}{0.9} % 降低行间距，缩小表格高度
% \begin{tblr}{
%   cells = {c},
%   hline{1-2,11} = {-}{},
% }
% Work                & Method           & Datasets  & Number of Subjects & ID Accuracy (\%) \\
% Chen \textit{et al.}\cite{chen}   & SVM              & MIT-BIH   & 48                 & 80               \\
% Zhao \textit{et al.}\cite{zhao_eemd} & EEMD+K-NN & MIT-BIH & 90&95 \\
% Zhang \textit{et al.}\cite{zhang} & Multiresolution 1D-CNN & MIT-BIH & 47&93.50 \\
% Bento \textit{et al.}\cite{bento}        & Spectrogram+CNN  & ECG-ID    & 90                 & 73.54            \\
% Patro \textit{et al.}\cite{patro_ecg_2020}        & Lasso+K-NN  & ECG-ID    & 20              & 99.14           \\
% Fatimah \textit{et al.}\cite{Fatimah}        & Fourier decomposition+Random Forest  & ECG-ID    & 89              & 97.92          \\
% Pourbabaee \textit{et al.}\cite{Pourbabaee_Howe-Patterson_Reiher_Benard_2018} & 1D-CNN & Self-built & 33 & 95.25\\
% Zhao \textit{et al.}\cite{Zhao} & 1D-CNN+Multisimilarity Loss & Self-built & 20&99.86  \\
% The proposed method & DT4ECG & Self-built & 15 & 99.12  
% \end{tblr}
% }
% \end{table*}
% Please add the following required packages to your document preamble:
% \usepackage{booktabs}
% \usepackage{multirow}
% Please add the following required packages to your document preamble:
% \usepackage{booktabs}
\begin{table*}[ht]
\caption{Descriptions and Results of ECG-based Identity Recognition and HAR in SOTA Approaches}
\label{tab:performance_comparison_SOTA}
\centering
\scalebox{1}{
\begin{tabular}{@{}ccccc@{}}
\toprule
Work (ID)                                                                                                           & Method                        & Datasets                                   & Subject & ID Accuracy (\%)  \\ \midrule
Chen \textit{et al.}\cite{chen}                                                   & SVM                           & MIT-BIH                                    & 48      & 80                \\
Zhao \textit{et al.}\cite{zhao_eemd}                                             & EEMD+K-NN                     & MIT-BIH                                    & 90      & 95                \\
Patro \textit{et al.}\cite{patro_ecg_2020}                                      & LASSO+K-NN                    & ECG-ID                                     & 20      & 99.14             \\
Bento \textit{et al.}\cite{bento}                                                 & Spectrogram+CNN               & ECG-ID                                     & 90      & 73.54             \\
Pourbabaee \textit{et al.}\cite{Pourbabaee_Howe-Patterson_Reiher_Benard_2018} & 1D-CNN                        & Self-built                                 & 33      & 95.25             \\
Zhao \textit{et al.}\cite{Zhao}                                                   & 1D-CNN+Multidiscilpinary Loss & Self-built                                 & 20      & 99.86             \\
\textbf{The proposed method}                                                                                                 & \textbf{1D-ResNet+SCA}                 & \textbf{Self-built}                                 & \textbf{15}      & \textbf{99.12}             \\ \midrule\midrule
Work (HAR)                                                                                                          & Method                        & Datasets                                   & Class   & HAR Accuracy (\%) \\ \midrule
Arani \textit{et al.}\cite{Arani_ECG_HAR}                                       & Random Forest                 & PPG-DaLiA\cite{PPG-DaLiA} & 5       & 88.44             \\
O'Halloran \textit{et al.}\cite{OHalloran}                                                                                                   & MLP                           & MHEALTH\cite{mhealth_319}                                    & 12      & 90.55             \\
Yatabaz et al.\cite{Yatbaz}                                                                                                      & CNN                           & MHEALTH\cite{mhealth_319}                                    & 12      & 93.05             \\
Yun \textit{et al.}\cite{Yun_HAR}                                                & 1D-CNN                        & Self-built                                 & 5       & 82.86             \\
\textbf{The proposed method}                                                                                                 & \textbf{1D-ResNet+SCA}                 & \textbf{Self-built}                                 & \textbf{3}       & \textbf{90.11}             \\ \bottomrule
\end{tabular}
}
\vspace{-1em}
\end{table*}

The experiments were conducted using the PyTorch framework with the following hyperparameters:

\begin{itemize}
    \item \textbf{Learning Rate}: 0.001
    \item \textbf{Optimizer}: Adam optimizer was used for all models.
    \item \textbf{Batch Size}: 16
    \item \textbf{Epochs}: 50 epochs for all models.
    \item \textbf{Loss Function}: Cross-entropy loss and GradNorm were used for all classification tasks.
    \item \textbf{Data Split}: The dataset was divided into a training set and a testing set with a ratio of 7:3.
    \item \textbf{Activation Function}: ReLU
    \item \textbf{Rate of Loss Adjustment \textbf{$\alpha$}}: 2
\end{itemize}

All models were trained under identical conditions to ensure a fair comparison.

\subsection{Evaluation Metrics}

We evaluate the models using the following four commonly used metrics in classification tasks:

\begin{equation}
Accuracy = \frac{TP + TN}{TP + TN + FP + FN}
\end{equation}

\begin{equation}
Precision = \frac{TP}{TP + FP}
\end{equation}

\begin{equation}
    Recall = \frac{TP}{TP + FN}
\end{equation}

\begin{equation}
    F1{-}score = 2 \times \frac{Precision \times Recall}{Precision + Recall}
\end{equation}
where TP, TN, FP, and FN refer to true positive, true negative, false positive, and false negative, respectively. These metrics provide a comprehensive evaluation of each model's performance in both classification accuracy and its ability to correctly identify relevant samples.

\subsection{Results and Comparisons}
The performance of our proposed model is summarized in Table \ref{tab:performance_comparison}. From the results, we observe that the proposed method achieves high performance in both person identification and HAR tasks. In the person identification task, it achieves 99.12\% accuracy, 99.14\% precision, 99.12\% recall, and 99.12\% F1-score on our self-built dataset of 15 subjects. As shown in Table \ref{tab:performance_comparison_SOTA}, the performance of our model in ECG-based person identification is comparable to that of SOTA methods reported in the literature, despite differences in datasets and experimental setups.

In addition to its high accuracy in person identification, the proposed method extends its capabilities to the HAR task, achieving an impressive 90.11\% accuracy on our self-built dataset with three activity classes. This demonstrates that the incorporation of MTL does not compromise the performance of either task. Instead, it enables the model to learn complementary representations that benefit both identity and activity classification tasks.
% Please add the following required packages to your document preamble:
% \usepackage{booktabs}
% \usepackage{multirow}

\subsection{Ablation Studies}
To further verify the effectiveness of the proposed SCA module and GradNorm loss, ablation studies are conducted in this section.
\subsubsection{Ablation Study on Attention Mechanisms}
To verify the effectiveness and generality of the proposed SCA mechanism, impacts of attention modules on the performance of different model backbones are investigated. The comparison of attention mechanisms is conducted among classical attention modules, including Squeeze-and-Excitation (SE)\cite{SE}, Multi-Head Attention (MHA)\cite{vaswani2023attentionneed}, Convolutional Block Attention Module (CBAM)\cite{CBAM}. All attention mechanisms are applied on 5 classical 1D-CNN backbones, including 1D-CNN, 1D-EfficientNet, 1D-MobileNet, 1D-DenseNet and 1D-ResNet. The results of the experiments are summarized in Table~\ref{tab:abl_attention_modules}. From the results, in the ID recognition task, all models and attention mechanisms achieved very high accuracy, with relatively small differences between them. For instance, the baseline 1D-CNN reached an ID accuracy of 98.12\%, and all attention mechanisms (SE, MHA, CBAM, and SCA) provided similar performance, with the highest being 99.00\% for SCA in the 1D-CNN backbone. This suggests that while attention mechanisms provide some enhancement, the ID recognition task is relatively straightforward, and thus the differences in performance across attention modules are minimal. 

In contrast to the identity recognition task, the activity classification task revealed more substantial differences in performance across attention mechanisms. In the 1D-CNN backbone, SCA achieved the best activity accuracy of 76.47\%, which is a 2.00\% improvement over SE (74.47\%), the second-best performer. For the 1D-EfficientNet backbone, SCA again achieved the highest activity accuracy of 86.36\%, outperforming SE (84.23\%) and CBAM (82.60\%). The 1D-MobileNet model, which had the highest ID accuracy with CBAM (99.25\%), saw SCA achieve the best activity accuracy (86.73\%), again outperforming CBAM (83.35\%). In the case of 1D-DenseNet, while SCA's ID accuracy was slightly lower (98.37\%), it provided the best activity accuracy (82.35\%). Finally, in 1D-ResNet, SCA achieved the highest activity accuracy of 90.11\%, outshining CBAM by 3.50\% (CBAM had 86.61\%).

Across all five tested backbones, the SCA mechanism consistently provided the best performance in the activity classification task. This indicates that SCA is particularly effective in capturing the temporal and contextual relationships essential for activity classification, which is typically more complex than ID recognition. The SCA's ability to enhance activity classification is clear from the improvement in accuracy across all model architectures, demonstrating its robustness and generality.
\begin{table}[]
\caption{Performance comparison of different attention modules in different backbones. The best performing metrics are highlighted in \textcolor{red}{\textbf{red}}, and the second-best in  \textcolor{blue}{\textbf{blue}}}
\label{tab:abl_attention_modules}
\begin{tabular}{@{}cclllcc@{}}
\toprule
Backbone                         & \multicolumn{4}{c}{\begin{tabular}[c]{@{}c@{}}Attention\\ Mechanism\end{tabular}} & ID Accuracy (\%) & HAR Accuracy (\%) \\ \midrule
\multirow{5}{*}{1D-CNN}          & \multicolumn{4}{c}{\XSolid}                                                          & 98.12            & 71.46               \\
                                 & \multicolumn{4}{c}{SE}                                                            & 98.62            & \textcolor{blue}{\textbf{74.47}}               \\
                                 & \multicolumn{4}{c}{MHA}                                                           & 98.37            & 72.97               \\
                                 & \multicolumn{4}{c}{CBAM}                                                          & \textcolor{blue}{\textbf{98.87}}            & 73.47               \\
                                 & \multicolumn{4}{c}{SCA}                                                           & \textcolor{red}{\textbf{99.00}}   & \textcolor{red}{\textbf{76.47}}      \\ \midrule
\multirow{5}{*}{1D-EfficientNet} & \multicolumn{4}{c}{\XSolid}                                                          & \textcolor{blue}{\textbf{98.75}}            & 82.10                \\
                                 & \multicolumn{4}{c}{SE}                                                            & 98.75            & \textcolor{blue}{\textbf{84.23}}               \\
                                 & \multicolumn{4}{c}{MHA}                                                           & 97.87            & 80.73               \\
                                 & \multicolumn{4}{c}{CBAM}                                                          & 98.75            & 82.60               \\
                                 & \multicolumn{4}{c}{SCA}                                                           & \textcolor{red}{\textbf{99.12}}   & \textcolor{red}{\textbf{86.36}}      \\ \midrule
\multirow{5}{*}{1D-MobileNet}    & \multicolumn{4}{c}{\XSolid}                                                          & 99.00            & 79.22\\
                                 & \multicolumn{4}{c}{SE}                                                            & 99.00            & 83.40               \\
                                 & \multicolumn{4}{c}{MHA}                                                           & \textcolor{blue}{\textbf{99.12}}            & \textcolor{blue}{\textbf{86.11}}               \\
                                 & \multicolumn{4}{c}{CBAM}                                                          & \textcolor{red}{\textbf{99.25}}   & 83.35               \\
                                 & \multicolumn{4}{c}{SCA}                                                           & 99.00            & \textcolor{red}{\textbf{86.73}}      \\ \midrule
\multirow{5}{*}{1D-DenseNet}     & \multicolumn{4}{c}{\XSolid}                                                          & \textcolor{red}{\textbf{99.00}}   & 79.72               \\
                                 & \multicolumn{4}{c}{SE}                                                            & \textcolor{blue}{\textbf{98.75}}            & \textcolor{blue}{\textbf{80.35}}               \\
                                 & \multicolumn{4}{c}{MHA}                                                           & 98.37            & 79.47               \\
                                 & \multicolumn{4}{c}{CBAM}                                                          & 98.75            & 77.85               \\
                                 & \multicolumn{4}{c}{SCA}                                                           & 98.37            & \textcolor{red}{\textbf{82.35}}      \\ \midrule
\multirow{5}{*}{1D-ResNet}       & \multicolumn{4}{c}{\XSolid}                                                          & 98.37            & 83.98               \\
                                 & \multicolumn{4}{c}{SE}                                                            & 98.62            & 84.48               \\
                                 & \multicolumn{4}{c}{MHA}                                                           & 98.50            & 82.60               \\
                                 & \multicolumn{4}{c}{CBAM}                                                          & \textcolor{blue}{\textbf{99.00}}            & \textcolor{blue}{\textbf{86.61}}               \\
                                 & \multicolumn{4}{c}{SCA}                                                           & \textcolor{red}{\textbf{99.12}}   & \textcolor{red}{\textbf{90.11}}      \\ \bottomrule
\end{tabular}
\end{table}

\subsubsection{Ablation Study on the Impact of GradNorm on Model Convergence}
\begin{figure}[htbp]
    \centering

    \subfloat[]{%
        \includegraphics[width=0.55\columnwidth]{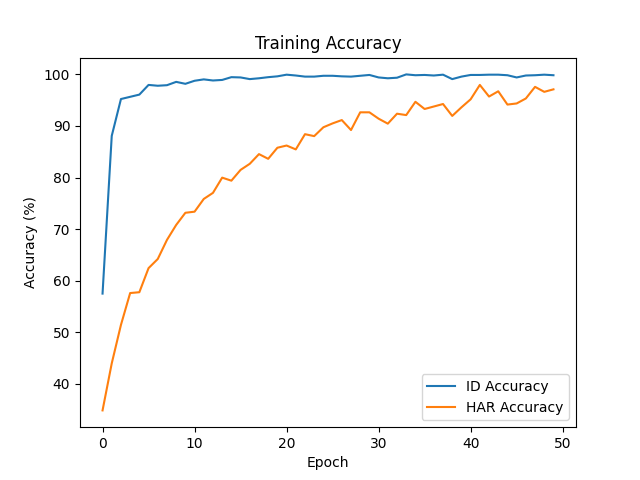}
        \label{fig:curve_without}
    }
    \subfloat[]{%
        \includegraphics[width=0.55\columnwidth]{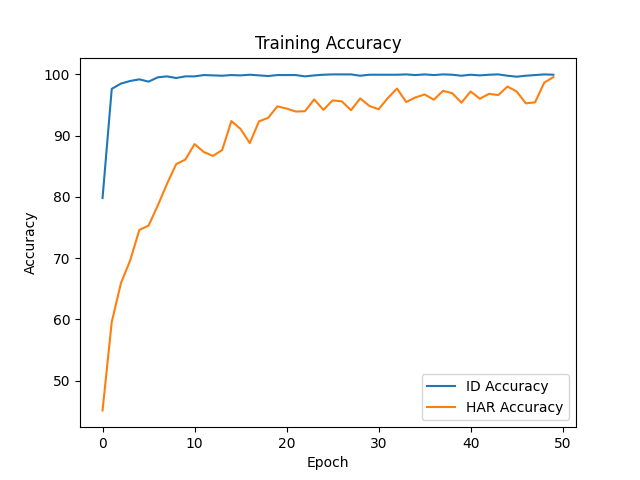}
        \label{fig:curve_with}
    }
    
    \caption{Training accuracy curve in 50 epochs. (a) Network without GradNorm (b) network with GradNorm}
    \label{fig:training curve}
\end{figure}

In this section, we perform an ablation study to evaluate the impact of GradNorm on the model's convergence speed and task balance. First, we present the training accuracy curves of the network with and without GradNorm under the same learning rate of 0.001 over 50 epochs in Fig.~\ref{fig:training curve}. The training accuracy curve in Fig.~\ref{fig:curve_without} indicate that, during the training of the network without GradNorm, the person classification task converges around 10 epochs, while the activity classification task progresses much slower, only converging after approximately 40 epochs. This imbalance in convergence speed can lead to inefficient utilization of computational resources.
From Fig.~\ref{fig:curve_with}, after the utilization of GradNorm mechanism, the training process of activity classification task accelerates, converging at around 35 epochs.

On the other hand, Fig.~\ref{fig:curve_with} demonstrates that with the application of the GradNorm mechanism, the training process of the activity classification task is accelerated, reaching convergence at around 35 epochs. This improved task balance enables the model to more efficiently allocate resources, optimizing both tasks simultaneously. 

\FloatBarrier

\begin{figure}[htbp]
    \centering

    \subfloat[]{%
        \includegraphics[width=0.5\columnwidth]{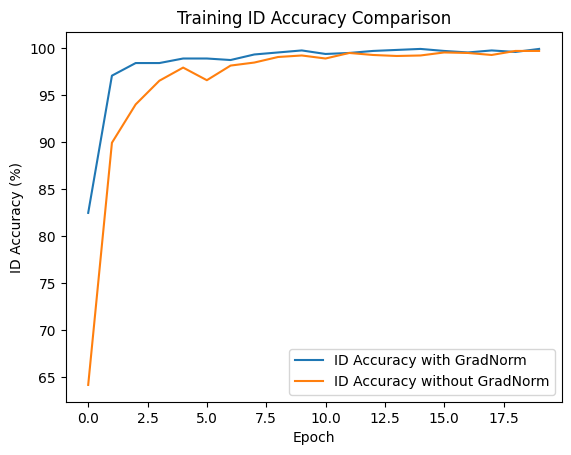}
        \label{fig:grad_person}
    }
    \subfloat[]{%
        \includegraphics[width=0.56\columnwidth]{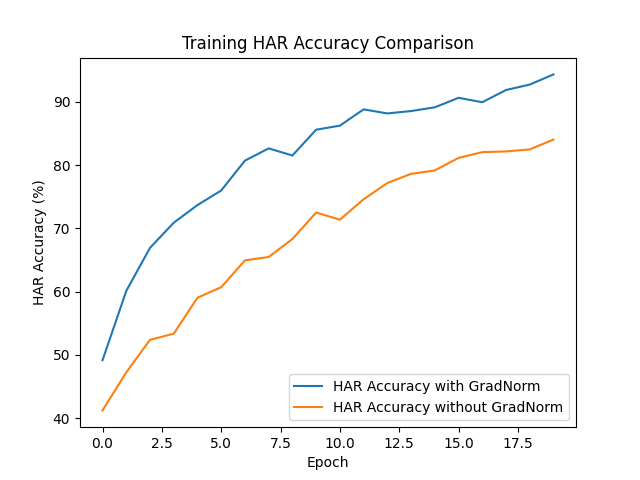}
        \label{fig:grad_state}
    }
    
    \caption{Training accuracy curve with and without GradNorm in 20 epochs. (a) ID classification (b) HAR classification}
    \label{fig:accuracy_curve_ablation}
\end{figure}

To further quantify the impact of GradNorm on training speed and test set performance, we conduct an additional experiment using a relatively smaller number of epochs (20 epochs). This allows us to observe GradNorm's influence during the early stages of MTL. The Adam optimizer with a learning rate of 0.001 is used for training the models. The accuracy curves for person classification and activity classification are shown in Fig.~\ref{fig:accuracy_curve_ablation}.

\FloatBarrier

In the training process for the person classification task, the application of GradNorm slightly accelerated the convergence speed, allowing the model to achieve optimal accuracy within just 3 epochs. However, the impact of GradNorm on the activity classification task is notably more pronounced and can be clearly analyzed through the data.

From the data presented in Fig. \ref{fig:grad_state}, it's evident that the network utilizing GradNorm consistently achieves approximately 10\% higher accuracy in each epoch compared to the network without GradNorm. This persistent performance enhancement indicates that GradNorm effectively adjusts the loss in the activity classification task. This not only accelerates convergence but also significantly improves accuracy in small number of epochs. Within 20 epochs, the network employing GradNorm surpasses 90\% accuracy on the training set, demonstrating a noticeably rapid convergence characteristic.
% Please add the following required packages to your document preamble:

% \usepackage{booktabs}
% Please add the following required packages to your document preamble:
% \usepackage{booktabs}
\begin{table}[htbp]
\caption{Performance comparison of the model with and without GradNorm after training 20 epochs}
\label{tab:gradnorm_results}
\centering % 使整个表格居中
\resizebox{\columnwidth}{!}{\begin{tabular}{@{}cccc@{}}
\toprule
Model                          & GradNorm                    & ID Accuracy (\%) & HAR Accuracy (\%) \\ \midrule
\multirow{2}{*}{1D-ResNet+SCA} & \Checkmark   & \textbf{98.75}               & \textbf{84.11}          \\
                               & \XSolidBrush & 95.13               & 78.22          \\ \bottomrule
\end{tabular}}
\end{table}

Further analysis is supported by the experimental results summarized in Table \ref{tab:gradnorm_results}, which compares the test set accuracy of models with and without GradNorm after 20 epochs of training. In the activity classification task, the model with GradNorm achieved an accuracy of 84.11\%, whereas the model without GradNorm reached only 78.22\%. This improvement suggests that within a relatively small number of epochs, the GradNorm mechanism effectively accelerates the training of more challenging task, thereby enhancing performance on the test set.
\section{Conclusion}
\label{sec::conclusion}
In this paper, we presented DT4ECG, a dual-task learning framework designed for ECG-based human identity recognition and activity activity detection. By leveraging the strengths of a CNN backbone integrated with 1D residual modules, our framework effectively extracts discriminative features from ECG signals. The introduction of the SCA mechanism allows the model to enhance feature representation by focusing on informative signal aspects across both temporal and channel dimensions.

Moreover, the incorporation of GradNorm dynamically balances the learning objectives in multi-task settings, ensuring improved convergence rates and enhanced performance for both identity and activity classification tasks. Our experimental results highlight the superiority of the proposed framework, achieving accuracy rates of 99.12\% in person classification and 90.11\% in activity classification on our custom-built dataset. These results underscore the potential of using ECG signals not only as a biometric identifier but also as a tool for activity activity detection.

The proposed DT4ECG framework enables efficient secure person authentication and robust real-time activity classification. This solution simplifies the integration of identity verification and activity monitoring into a single system, making it ideal for applications in personalized health tracking, context-aware services, and secure authentication in wearable devices.

%\begin{thebibliography}{99}
\bibliographystyle{IEEEtran}
\bibliography{IEEEabrv,Ref}

%\end{thebibliography}

\end{document}